\documentclass[10pt,twocolumn,pagenumbers]{IEEEtran}

\makeatletter
\let\NAT@parse\undefined
\makeatother

\usepackage{amsmath}
\usepackage{amssymb}
\usepackage{caption3}
\usepackage{xy}
\usepackage[vflt]{floatflt}
\usepackage{multicol}
\usepackage[dvips]{graphicx}
\DeclareGraphicsExtensions{.eps}

\usepackage{hyperref}

\newfont{\hiera}{cmsy10 scaled 2488} 
\newfont{\hierb}{cmsy10 scaled 1728}
\newfont{\hierc}{cmsy10 scaled 1200}

\newcommand{\Bigast}{
\mathop{\vphantom{\sum}\lower2.5pt\hbox{\hiera\char3}}}%

\newcommand{\Bigtimes}{
\mathop{\vphantom{\sum}\lower2.5pt\hbox{\hiera\char2}}}%

\def\BibTeX{{\rm B\kern-.05em{\sc i\kern-.025em b}\kern-.08em
    T\kern-.1667em\lower.7ex\hbox{E}\kern-.125emX}}

\begin{document}

\title{\Large ``On the engineers' new toolbox''\\
or\\
Analog Circuit Design, using Symbolic Analysis, Computer Algebra, and Elementary Network Transformations\\
{\large (2011 Re-Release)}
}

\author{{\large Eberhard H.-A.\ Gerbracht
\thanks{This article first appeared in: SM$^2$ACD'08 Proceedings of the Xth International Workshop on Symbolic and Numerical Methods, Modeling and Applications to Circuit Design, Erfurt, Germany, October 07 to 08, 2008; pp.~ 127-134. Due to the low distribution of these proceedings, the author has decided to make the article available to a larger audience through the arXiv.}
\thanks{The author's current (as of December~31st, 2010) address is Bismarckstra\ss e 20, D-38518 Gifhorn, Germany. Current e-mail: \tt{e.gerbracht@web.de}}}}
 
\maketitle
\thispagestyle{empty}

\bigskip

\begin{abstract}
\noindent
{\small \bf
In this paper, by way of three examples -- a fourth order low pass active RC filter, a rudimentary BJT amplifier, and an LC ladder -- we show, how the algebraic capabilities of modern computer algebra systems can, or in the last example, might be brought to use in the task of designing analog circuits.
}
\end{abstract}

{\small
\smallskip

\smallskip

}

\section{Introduction}

For those about to embark on the study of this paper, let us start with a warning: This article is not a report on ground-breaking results. Neither will we present spanking new circuits, ripe for patent. Rather in spirit of SM$^2$ACD's explicit policy to be ``a forum for the discussion of new ideas and methodologies'', we would like to advocate the \hbox{(re)}\-introduction\footnote{%
{\bf Re}introduction, because some symbolic analysis tools were and still are closely associated to existing computer algebra systems, although nowadays they do not make much use of the advanced algebraic capabilities (it is debatable if they ever did), but rely more on their numeric proficiencies.} of computer algebra into the design process of analog circuits. In order to do this, we present some examples, where computer algebra (together with numerics) was used in the design, especially the sizing of the given circuits.

These examples do not represent a systematic approach, but they should be considered as tentative steps towards getting to know the capabilities (and some of the shortcomings) of computer algebra systems (CAS), and making these tools bring forth fruit in the daily work of electrical engineers. 

\section{The general ansatz}

\noindent 
One of the principal problems in analog circuit design\footnote{See e.g.\ \cite{GielenSansen}.} is to determine element values in a given circuit in such a way that the resulting network function is in accordance with some previously given performance specifications. In more mathematical terms: suppose the network function is demanded to be a rational function of form
\begin{equation}
f_{des}(s)  =  \frac{a_0+a_1s+\dots+a_m s^m}{b_0+b_1s+\dots+b_{n}s^{n}},
\end{equation}
\noindent
with the numbers $m, n,$ $a_1,\dots,a_{m},$ $b_1,\dots,b_n$ determined by the initial specifications. 

Then we need to find an analog circuit $\Gamma$ built from elements, which themselves are described by parameters $p_1,\dots,p_k$, such that its network function
\begin{equation}
H_\Gamma(s)  =  
\frac{\sum_{\mu=0}^{m}A_\mu(p_1,\dots,p_k) s^i}
{\sum_{\nu=0}^{n}B_\nu(p_1,\dots,p_k) s^i},
\end{equation}
\noindent
where the terms $A_\mu$ and $B_\nu$ are polynomials in $p_1,\dots,p_k,$
satisfies
\begin{equation}
H_\Gamma(s) = f_{des}(s).
\end{equation}
\noindent
In other words: at the end of the day, many design problems result in a system of nonlinear equations
\begin{equation}
\begin{aligned}
A_\mu(p_1,\dots,p_k) - a_\mu & = & 0,\\
B_\nu(p_1,\dots,p_k) - b_\nu & = & 0,
\label{DesignProblem}
\end{aligned}
\end{equation}

\noindent
$1\le \mu\le m,$ and $1\le \nu \le n$ for variables $p_1,\dots,p_k.$

\noindent
If indeed the terms $A_\mu, B_\nu$ are polynomials\footnote{Thus for example we do not talk about the problem of biasing a transistor, since then we would have to solve equations containing the exponential function (as in the Ebers-Moll model of the BJT), or equations which contain roots, and moreover are piecewisely defined (as for FETs).} with coefficients in some number field, 
the set of solutions, other\-wise also known as the zero set of (\ref{DesignProblem}), 
forms what is called an {\sl affine algebraic variety}, an object which belongs to the mathematical subdiscipline  
 of {\sl Algebraic Geometry.}
Thus when trying to solve these equations, we could profit both from the centuries of experience, mathematicians have gained in this particular field, and from
the more recent results and algorithmic instruments developed
-- the concept of {\sl Gr\"obner bases,} as well as the {\sl Buchberger algorithm} and its ``relatives'' which are realized  in nearly all of the current computer algebra systems. 
The result would be exact (``symbolic'') descriptions of the sought after solution sets, from which we could deduce numeric results afterwards. 

\noindent
Now, when we take a look at the advances in symbolic methods applied to circuit design, we can make a peculiar observation: nowadays it has become nearly routine to calculate any given network function as a multivariable rational function of the complex frequency $s$ and the symbolic circuit elements (even though there is still the problem of the combinatorial explosion for large circuits). However, when it comes to finding solutions for the nonlinear equations arising from the task of sizing a particular circuit, even if it is relatively small, researchers, as well as designers, fall back upon numerical means and various approximative methods with their various drawbacks.


\section{An Example - Sizing an active 4th order low pass RC filter}

In section 6.3.3.\ of \cite{HieuDiss} the following filter circuit was presented:

\begin{figure}[!ht]
\begin{center}
\includegraphics[width=\linewidth,clip,keepaspectratio]{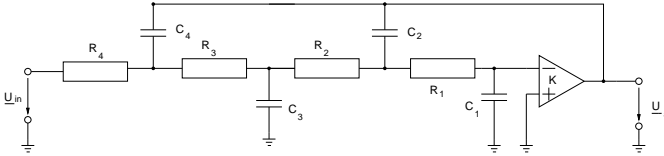}
\parbox{.8\linewidth}{
\caption{Circuit topology for fourth order low pass RC filter}
}
\end{center}
\end{figure}

\noindent
 The aim there was to construct a fourth order Butterworth filter  with this topology. 

Using one of the available symbolic analysis programs%
\footnote{For the analysis of this and all other circuits within this paper, we use SapWin 3.0, because of its affordability and its easy availability through the world wide web \cite{SapWin}.}, we see that the circuits symbolic transfer function is given by
\begin{equation}
H(s):=\frac{\underline{U}_a}{\underline{U}_{in}}=\frac{K}{D(s)}
\end{equation}
\noindent
with
$
D(s)=1+a_1 s+a_2 s^2+ a_3 s^3 + a_4 s^4
$
 being a polynomial, and

\vskip -2em
\begin{eqnarray}
  a_1 &=&
\scriptstyle
((1-K)(C_4+C_2)+C_3+C_1) R_4 + ((1-K)C_2+C_3+C_1) R_3 + \nonumber\\
&&
\scriptstyle
((1-K)C_2+C_1) R_2 + C_1 R_1,\nonumber\\
  a_2 &=& 
\scriptstyle
C_2 C_1 R_1 R_2
 + (C_3 + C_2) C_1 R_1 R_3
 + (C_4 + C_3 + C_2) C_1 R_1 R_4  +\nonumber\\
&&
\scriptstyle
C_3 ((1 - K) C_2 + C_1) R_2 R_3
 +
((C_4+C_3)((1 - K) C_2+C_1) R_2 R_4 +\nonumber\\
&&
\scriptstyle
 C_4 (C_3 + (1 - K) C_2 + C_1) R_3 R_4,\label{4FilterGlg}
\\
  a_3 &=& 
\scriptstyle
C_3 C_2 C_1 R_1 R_2 (R_3 + R_4 )
 + C_4 C_2 C_1 R_1 (R_2 + R_3) R_4 
 +\nonumber\\
&&
\scriptstyle
 C_4 C_3 C_1 (R_1 + R_2) R_3 R_4 
 + (1 - K) C_4 C_3 C_2 R_2 R_3 R_4,
\nonumber\\
  a_4 &=& 
\scriptstyle C_4 C_3 C_2 C_1 R_1 R_2 R_3 R_4.\nonumber
\end{eqnarray} 

\noindent
In order to specify element values in such a way that $H(s)$ becomes the transfer function of a fourth order Butterworth filter, we can proceed from these data by matching the coefficients with those numerical values, which can be found in any table  for filter design, and then try to solve the resulting equations by numerical means. Since we only have four equations for nine unknowns, there is the need to reduce the existing degrees of freedom by more or less arbitrarily attaching numerical values to some of the element values. In fact, that is what was done in \cite{HieuDiss}, where $K$ was set equal to $2$ and all the $C_i$ were set to $1$ (all element values in this example are considered to be normalized); afterwards some variant of the Newton-Raphson method was used, to deduce one single solution.

\noindent
The disadvantages of such an approach are obvious: Using rounded values (from precalculated tables) introduces errors right from the start.  Both, the assignment of (quite arbitrary) numbers to arbitrarily chosen parameters, and the choice of a starting vector for the Newton-Raphson method might hinder from finding a solution, or even worse, might make a solution impossible\footnote{E.g.\ this is the case, when we use the preassigned values from \cite{HieuDiss}, and try to find element values, such that the above circuit attains the characteristics of a fourth order Chebyshev filter.}       

Furthermore, questions like ``Do alternatives exist? How many are there?'', cannot be answered either, and changing any of the numerical parameters forces us to redo most of the calculations.

When we tried to solve this design task, first of all we started with a slightly different approach, using the Feldtkeller equation 
\begin{equation}
D(s)\cdot D(-s) = 1 + s^8
\end{equation}
for the special case of the fourth order Butterworth filter. Thus we got rid of the precalculated approximate values for the $a_i,$ by using instead the defining algebraic equations given by matching the coefficients from
\begin{eqnarray*}
D(s)\cdot D(-s)&=& 1 + (2a_2 - a_1^2) s^2 
+ (a_2^2 - 2 a_1 a_3 + 2 a_4) s^4\\
&&\phantom{1}+ (2 a_2 a_4 - a_3^2) s^6 
+ a_4^2 s^8
\end{eqnarray*}
with $1+s^8.$ Consequently we enlarged the set of equations (\ref{4FilterGlg}) by adding
\begin{equation}
2a_2 - a_1^2 = 0,\, a_2^2 - 2 a_1 a_3 + 2 a_4=0,\, 2 a_2 a_4 - a_3^2=0,\, a_4=1
\label{BW}
\end{equation}
to it, and regarding $a_1,\dots, a_4$ as further unknowns.

Even though in this paper we propose the usage of the algebraic capabilities of a CAS, at this point we used Mathematica (Version 6.0.1.0) as a numerical engine -- not in order to approximate one solution, but to find approximates for {\bf all} solutions. Moreover as was done in \cite{HieuDiss}, we also fix $K=2,$ and $C_i=1$ for $1\le i\le 4.$ Since obviously the enlargement of the set of equations results in a bigger set of solutions, we were left with the final task of ``post-processing'' this zero set, by singling out those particular solutions for which $a_1,\dots,a_4$ determine a polynomial $D(s)$, all of whose zeroes have negative real part (= filter stability), and such that all the $R_i$ are {\sl non-negative} (= realizability by classic elements). Indeed, we observe that from all the possible solutions calculated, only nine satisfy the restrictions concerning the $a_i,$ and again only one of these, 
\begin{eqnarray*}
K &=& 2,\ C_1=C_2=C_3=C_4=1,\\
R_1 &\approx& 0.133933818297194652631087580090,\\
R_2 &\approx& 3.893036697318392402871746149006,\\
R_3 &\approx& 2.479192111455558403082198766696,\\
R_4 &\approx& 0.773590398536329977043175927841.
\end{eqnarray*}
results in the $R_i$ all being positive.

So, where does computer algebra enter this picture?

First of all Mathematica's {\tt NSolve}-procedure, which we used in the above task, at its core heavily depends on algebraic algorithms: the fact that it is able to calculate {\sl all} solutions stems from the algebraic variety, which was defined by (\ref{4FilterGlg}) and (\ref{BW}), being {\sl zero dimensional}. That is, it consists of finitely many points.
Again we can be sure of this fact, only because there is an algorithm -- the {\sl Buchberger algorithm\footnote{%
The Buchberger algorithm can be seen as a generalization of both the Euclidean algorithm (to the multidimensional case) and the Gau\ss\ algorithm (to polynomial equations of degree $> 1$).}} -- which is used by Mathematica as a subprocedure of {\tt NSolve} and which for any system of algebraic equations produces particularly well-behaved equivalent systems of equations -- so-called {\sl Gr\"obner bases} -- which allow much more insight into the properties of the zero set, and may be used for a subsequent elimination of variables\footnote{In most cases, especially if the zero set is zero dimensional, there exist Gr\"obner bases which are in {\sl triangular form}, a property which generalizes the analogous property of sets of linear equations, which can be brought to triangular form using the Gau\ss\ algorithm. For more details see \cite{CoxIdeals,CoxUsing}.}. 

Thus e.g., when we apply Mathematica's {\tt GroebnerBasis}-command to (\ref{4FilterGlg}) and (\ref{BW}), with the values of $K,C_1,\dots,C_4$ fixed as above, and the variables ordered lexicographically by $a_1<\dots a_4<R_1<\dots<R_4$ (which amounts to saying that first $a_1$ is to be eliminated, then $a_2,$ etc.), then the resulting Gr\"obner basis contains the polynomial 

\medskip
\noindent
$
P =
 1 - 160 \, R_4^4 + 13872 \, R_4^8 - 788512 \, R_4^{12} + 31505120 \, R_4^{16}
$

\smallskip
\noindent
$ 
- 920274816 \, R_4^{20} + 20065991808 \, R_4^{24}
$

\smallskip\noindent 
$
- 328437088768 \, R_4^{28} + 4000414289152 \, R_4^{32}
$

\smallskip\noindent
$
 - 35535204282368 \, R_4^{36} + 223781766674432 \, R_4^{40}
$

\smallskip\noindent
$
 - 
 956822102532096 \, R_4^{44} + 2535921430958080 \, R_4^{48}
$

\smallskip\noindent
$
 - 
 3050522934050816 \, R_4^{52} - 2341746368053248 \, R_4^{56}
$

\smallskip\noindent
$
 + 
 10182414501412864 \, R_4^{60} - 1806331484831744 \, R_4^{64}
$

\smallskip\noindent 
$
- 
 13996296348106752 \, R_4^{68} + 2888816545234944 \, R_4^{72}
$

\medskip\noindent
in $R_4$ of degree $72,$ which implies that there are only finitely many possible values for $R_4.$ Furthermore this Gr\"obner basis contains polynomials $p_1(R_1,R_4),$ $p_2(R_2,R_4),$ and $p_3(R_3,R_4),$ each of which being linear in the respective $R_i,$ and otherwise consisting only of a high degree polynomial in $R_4.$ Since these polynomials as part of a Gr\"obner basis have to be zero, too, consequently each zero of $P$ determines precisely one solution vector $(R_1,\dots,R_4)$.  

Those solutions, which result in the additional specifications for the $a_i$ to be fulfilled, originate from a factor of the polynomial $P$ of degree $18,$ which can be calculated using Mathematica's {\tt Factor}-procedure, when we allow the coefficients to be sums of integers and integer multiples of $\sqrt{2}.$ This factor polynomial is
\vskip-1em
\begin{eqnarray*}
&&-10 + 7 \sqrt{2} + (-12 + 8 \sqrt{2}) R_4^2 
+ (280 - 196 \sqrt{2}) R_4^4\\
&&+ (520 - 352 \sqrt{2}) R_4^6 
+ (-2824 + 1900 \sqrt{2}) R_4^8\\ 
&&+ (-10816 + 
     6656 \sqrt{2}) R_4^{10}
+ (-13904 + 7368 \sqrt{2}) R_4^{12}\\ 
&&+ (-8288 + 
     5952 \sqrt{2}) R_4^{14}
+ (1280 + 6592 \sqrt{2}) R_4^{16}\\ 
&&+ 10368 R_4^{18}.
\end{eqnarray*}
\noindent
 Substituting $T=R_4^2,$ we see that its structure is effectively determined by a polynomial of degree $9,$ which results in the number of admissable solutions being nine.

To finish this section, let us remark that by modifying the Feldtkeller equation to 
\begin{equation}
D(s)\cdot D(-s) = 1 + T_4(s)\cdot T_4(-s),
\nonumber
\end{equation}    
where $T_4(s)$ denotes the fourth order Chebyshev polynomial,
and slightly changing the initial values for the $C_i,$ after some trial-and-error we were able to calculate the following set of parameter values, with which we can make the initially given circuit into a Chebyshev filter
\begin{eqnarray*}
K &=& 2,\ C_1 = 1,\ C_2 = 2,\ C_3 = 2,\ C_4 = 1,\\
R_1 &\approx& 0.263638090854794185461593787592,\\
R_2 &\approx& 0.624164765447879000316525786179,\\
R_3 &\approx& 2.645185226758545312744750592839,\\
R_4 &\approx& 3.249013399873649633437777451232.
\end{eqnarray*}

The question remains for each filter characteristic: which set of prescribed values for $K, C_1,\dots, C_4$ imply solutions for $R_1,\dots,R_4,$ which make sense in an electrical engineering point of view, i.e., for which positive $K, C_1,\dots, C_4$ do we get positive $R_1,\dots,R_4$? 
%
%
%
%
%
%
%
%
%
%
%
%
%
%
%
%
%

\section{Another Example: Sizing a BJT Amplifier with respect to break frequencies}

\begin{figure}[!h]
\begin{center}
\vskip -.4cm
\hskip-1em
{\includegraphics[width=1.1\linewidth,clip,keepaspectratio]{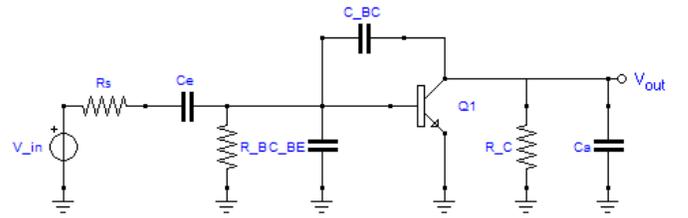}}
\vskip -0.4cm
\parbox{.7\linewidth}{
\caption{Rudimentary topology for a BJT amplifier}
}
\label{NaundorfPicture}
\end{center}
\vskip-2em
\end{figure}

From \cite{Naundorf} - a students' textbook in analog electronics - we take our next example: a rudimentary BJT amplifier circuit with only a few additional classic elements for biasing and sizing purposes (see figure (\ref{NaundorfPicture})).
The biasing resistances are given by
$
R_B = 310 \hbox{k}\Omega,
$ and 
$
R_C = 2 \hbox{k}\Omega.
$
The biased transistor is modelled by concrete h-parameters
\begin{equation}
H=
\begin{pmatrix}
h_{11} & h_{12}\\
h_{21} & h_{22}
\end{pmatrix},
=
\begin{pmatrix}
672 \Omega & 0\\
96 & 35\cdot 10\mu \hbox{S}
\end{pmatrix}
\end{equation}
and its capacitances $C_{BC} = 11 \hbox{pF}$, and $C_{BE} = 320 \hbox{pF}$.

\begin{figure}[h!]
\begin{center}
{\includegraphics[width=\linewidth,clip,keepaspectratio]{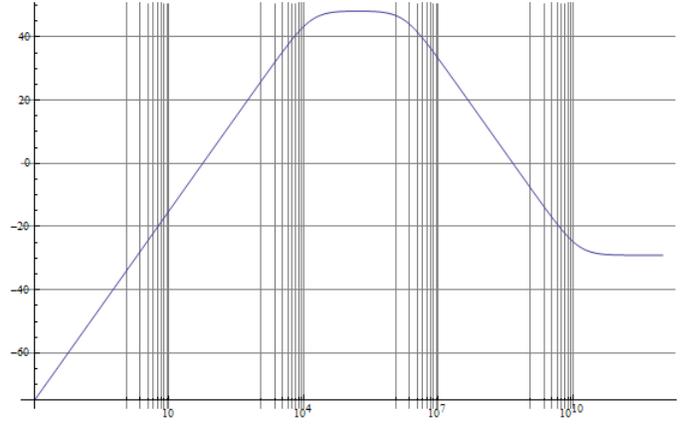}}
\parbox{.9\linewidth}{
\caption{Bode Plot of initially given BJT amplifier, according to (\ref{NaundorfBJT})}
\label{Naundorf1}
}
\vskip -0.4cm
\end{center}
\vskip -0.4cm
\end{figure}

With denoting its transfer function by $H(s) = \frac{N(s)}{D(s)},$  a full symbolic analysis with SapWin 3.0.\ gives
\begin{equation}
N(s)
=
C_e h_{21} R_B R_C \cdot s - C_{BC} C_e h_{11} R_B R_C\cdot s^2,
\nonumber
\end{equation}
and
$
D(s)= a_0 + a_1 s + a_2 s^2 + a_3 s^3,
$
with
\begin{eqnarray*}
a_0 & = &
\scriptstyle
 -h_{11} - R_B + h_{12} h_{21} R_C - h_{11} h_{22} R_C - 
 h_{22} R_B R_C,\\
a_1 
& = &
\scriptstyle
  - C_{BC} h_{11} R_B - C_{BE} h_{11} R_B - C_e h_{11} R_B - C_a h_{11} R_C\\
&&\scriptstyle
 - C_{BC} h_{11} R_C - C_a R_B R_C - C_{BC} R_B R_C + C_{BC} h_{12} R_B R_C\\ 
&&\scriptstyle
 - C_{BC} h_{21} R_B R_C + C_{BC} h_{12} h_{21} R_B R_C + C_{BE} h_{12} h_{21} R_B R_C\\
&&\scriptstyle
 + C_e h_{12} h_{21} R_B R_C - C_{BC} h_{11} h_{22} R_B R_C - C_{BE} h_{11} h_{22} R_B R_C\\
&&\scriptstyle
 - C_e h_{11} h_{22} R_B R_C - C_e h_{11} R_s - C_e R_B R_s + C_e h_{12} h_{21} R_C R_s\\
&&\scriptstyle
 - C_e h_{11} h_{22} R_C R_s - C_e h_{22} R_B R_C R_s,\\
a_2 & = &
\scriptstyle
-C_a C_{BC} h_{11} R_B R_C - C_a C_{BE} h_{11} R_B R_C - C_{BC} C_{BE} h_{11} R_B R_C\\
&&\scriptstyle
- C_a C_e h_{11} R_B R_C - C_{BC} C_e h_{11} R_B R_C - C_{BC} C_e h_{11} R_B R_s\\
&&\scriptstyle
 - C_{BE} C_e h_{11} R_B R_s - C_a C_e h_{11} R_C R_s - C_{BC} C_e h_{11} R_C R_s\\
&&\scriptstyle
 - C_a C_e R_B R_C R_s - C_{BC} C_e R_B R_C R_s + C_{BC} C_e h_{12} R_B R_C R_s\\
&&\scriptstyle
 - C_{BC} C_e h_{21} R_B R_C R_s + C_{BC} C_e h_{12} h_{21} R_B R_C R_s\\
&&\scriptstyle
 + C_{BE} C_e h_{12} h_{21} R_B R_C R_s - C_{BC} C_e h_{11} h_{22} R_B R_C R_s\\
&&\scriptstyle
 - C_{BE} C_e h_{11} h_{22} R_B R_C R_s,\\
a_3 &=&
\scriptstyle
-C_a C_{BC} C_e h_{11} R_B R_C R_s - C_a C_{BE} C_e h_{11} R_B R_C R_s\\
&&\scriptstyle
 - C_{BC} C_{BE} C_e h_{11} R_B R_C R_s.
\end{eqnarray*}
\noindent
With regard to the concretely given transistor parameters at the bias point,
the nominator and denominator of $H(s)$ become
\begin{equation}
N'(s)=
(290625000000 C_e s (-1000000000000 + 77 s)),
\nonumber
\end{equation}
and
\begin{eqnarray*}
D'(s)&=&
\phantom{+} 1623139843750000000 \\
&&+ 3590505265625\, s\\
&& + 3033906250000000000000\, C_a s\\
&& + 1088390625000000000000\, C_e s
\\
&& + 1623139843750000000\, C_e R_s s\\
&& + 7161\, s^2\\
&& + 673378125000000\, C_a s^2 + 22378125000000\, C_e s^2\\
&& + 2034375000000000000000000 C_a C_e s^2\\
&& + 3590505265625 C_e R_s s^2\\ 
&& + 3033906250000000000000 C_a C_e R_s s^2\\
&& + 7161 C_e R_s s^3 + 673378125000000 C_a C_e R_s s^3,
\end{eqnarray*}
\noindent
respectively. Thus, e.g.\ setting $R_s=0\Omega,$ $ C_a= 300\hbox{pF},$ and $C_b=0.1\mu\hbox{F},$ we get
\begin{equation}
H(s)=
\frac{\scriptstyle 465000 s (-1000000000000 + 77 s)}
{\scriptstyle 25970237500000000000 + 
 1813435834250000 s + 1015651791 s^2},
\label{NaundorfBJT}
\end{equation}
\noindent
the poles being (approximately\footnote{Clearly, due to the nominator being a quadratic function, we would also be able to calculate exact values.})
\begin{equation}
p_1\approx - 14437.758, \quad
\hbox{and}
\quad
p_2\approx - 1,771051.964.
\nonumber
\end{equation}
This results in the Bode plot shown in figure \ref{Naundorf1}.

\medskip

\noindent
We will now use this example as a problem in circuit design. Our task shall be: with the bias point remaining unchanged, find parameters $R_s, C_a,$ and $C_e$ in such a way that a new transfer function $\widetilde{H}(s)$ is achieved, which has the same zeroes as $H(s),$ and possesses two (new) poles $p_1$ and $p_2$.
Introducing a new parameter $k$ (since due to the simplicity of our design topology, we cannot specify the maximal attenuation of $\widetilde{H},$ as we will see below), we thus are asked to determine the parameters $k, R_s, C_a,$ and $C_e$ in such a way that
\begin{equation}
\frac{N'(s)}{D'(s)}
=
\widetilde{H}(s)
=
\frac{k\cdot s \cdot ( 77 s -1 000 000 000 000)}{(s-p_1)\cdot(s-p_2)}.
\end{equation}
This is equivalent to 
\begin{equation}
N'(s)\cdot (s-p_1)\cdot(s-p_2) - D'(s)\cdot k\cdot s \cdot ( 77 s -1000000000000)=0. 
\label{NaundorfDesign}
\end{equation}
Factoring out common terms on both sides of the $-$-sign, we are left with a third order polynomial in $s$ with coefficients being integer polynomials in the variables $\{k,R_s,C_a,C_e,p_1,p_2\}.$ To solve (\ref{NaundorfDesign}) we need to set these coefficients to zero, which consequently results in the following system of four algebraic equations\footnote{This is why we had to introduce the additional parameter $k,$ since we consider $p_1$ and $p_2$ to be fixed.} (the $i$-th equation resulting from the coefficient of $s^i$):
\vskip-1em
\begin{align}
0 &=
\phantom{+}
  1623139843750000000\, k - 290625000000\, C_e p_1 p_2,\nonumber\\
0 &=
\phantom{+} 3590505265625\, k + 3033906250000000000000\, C_a k\nonumber\\
& 
+ 1088390625000000000000\, C_e k + 290625000000\, C_e p_1\nonumber\\
& 
+ 290625000000\, C_e p_2 + 1623139843750000000\, C_e k R_s,\nonumber\\
0 &=
 -290625000000\, C_e + 7161\, k\label{NaundorfGlg}\\
&
+ 673378125000000\, C_a k + 22378125000000\, C_e k\nonumber\\
& 
+ 2034375000000000000000000\, C_a C_e k\nonumber\\
& 
+ 3590505265625\, C_e k R_s\nonumber\\
& 
+ 3033906250000000000000\, C_a C_e k R_s,\nonumber\\
0 &=
7161\, C_e k R_s + 673378125000000\, C_a C_e k R_s.\nonumber
\end{align}

\noindent
Alhough this system is nonlinear, using its procedure {\tt Solve}, Mathematica is able to calculate four symbolic solutions for $k, R_s, C_a,$ and $C_e$ as functions of $p_1$ and $p_2,$ voicing the explicit caveat: ``{\tt Equations may not give solutions for all "solve" variables.}''. Nevertheless, indeed there are only four different solutions.

The first is irrelevant for the concrete problem at hand from an electrical engineering point of view, since it is given by 
\begin{equation}
C_e=0\quad \hbox{and}\quad k=0,
\end{equation}
and thus results in a trivial transfer function.

The second one, though it leads to nontrivial terms for $C_e,k$ and $R_s,$ furthermore implies
\begin{equation}
C_a=-\frac{11}{1034375000000},
\label{mies}
\end{equation}
which again has to be considered absurd from an engineering point of view\footnote{Unless we allow for NICs (negative impedance converters) to be added to the initial circuit topology.}.

Finally, there are two solutions left which lead to two different valid sets of design parameters, both of them having in common that $R_s=0.$ 

Here we will not present the explicit formulas for $C_a, C_e, k,$ but we will give a numerical example, and we will further show, how the formulas for $C_a$ (and beyond this our claim, that there are only four solutions) can be derived with the help of Gr\"obner bases.  

First, for the numerical example let $p_1=-10,$ and $p_2=-1000.$ Then the two remaining approximate solutions of the above set of equations are  
\begin{equation}
(C_a, C_e, k) \approx (53.4989\mu\hbox{F}, 1.49102\mu\hbox{F}, 2.66969\cdot 10^{-9})
\label{zwei}
\end{equation}
and
\begin{equation}
(C_a, C_e, k) \approx (53.5000\mu\hbox{F}, 149.12881\mu\hbox{F}, 2.67017\cdot10^{-7}).
\label{drei}
\end{equation}

Indeed, the corresponding Bode plots for the respective circuits can be seen in figure 4.
\medskip
\begin{figure}[!ht]
\begin{center}
\includegraphics[width=\linewidth]{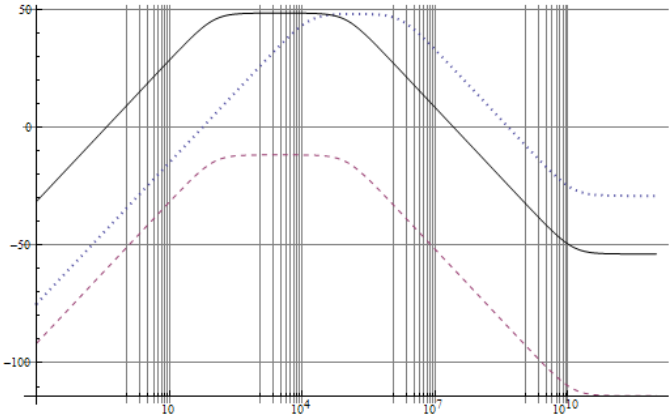
}
\vskip -0.4cm
\caption{Bode plot of BJT amplifier circuit with original parameters (dotted), parameters (\ref{zwei}) (dashed), and (\ref{drei}) (thick)}
\end{center}
\label{Bode3}
\end{figure}

Finally, applying Mathematica's {\tt GroebnerBasis}-comm\-and to the right-hand sides of equations (\ref{NaundorfGlg}), with the order of variables chosen to be $k < R_s < C_a < C_e,$ results in a Gr\"obner basis consisting of $11$ polynomials with exactly one of them containing only the variables $C_e, p_1, p_2.$ This particular polynomial is of degree four in the variable $C_e.$ Keeping in mind that the zero set of the original polynomials constituting (\ref{NaundorfGlg}) is equal to the zero set of its Gr\"obner basis, we thus see that there truly are only four possibilities for the values of $C_e.$

Factoring this particular polynomial (again with the help of Mathematica) we get the following factors:
\begin{align}
&C_e,\nonumber
\intertext{and}
&\phantom{ + }1182300335490970802500000000000000\nonumber\\
& - 2591834572818828634150000000\, p_1\nonumber\\
& - 2591834572818828634150000000\, p_2\\
& + 5681810493667293811609\, p_1\, p_2\nonumber\\
& + 1737403713997011459000000000000\, p_1\, p_2\, C_e, \nonumber
\intertext{and finally}
&100852820307500000000000000\nonumber\\
& - 22384370986950000000\, p_1\nonumber\\
& - 22384370986950000000\, p_2\nonumber\\ 
& + 49070938130697\, p_1\, p_2\nonumber\\
& - 67626498450000000000000000000\, p_1\, C_e\label{QQ}\\
& - 67626498450000000000000000000\, p_2\, C_e\nonumber\\ 
& + 163214120034000000000000\, p_1\, p_2\, C_e\nonumber\\ 
& + 45346707000000000000000000000000\, p_1\, p_2\, C_e^2.\nonumber
\end{align}    
Further analysis shows that $C_e=0$ results in $k=0,$ and the solution of the linear (with respect to $C_e$) equation implies the negative value of $C_a$ according to (\ref{mies}). Thus the two solutions, whose existence we claimed above, have their origin in the quadratic equation (\ref{QQ}). 

\section{What If equations become too complicated? An Outlook}

Symbolic methods always suffer from one major deficiency -- the ``combinatorial explosion'' of terms -- which most of the time results in both, a massive, nearly insatiable demand for space, and an extraordinary high amount of computing time. This observation also applies to all Gr\"obner basis methods:
It is known that by necessity the problem of finding a Gr\"obner basis in the general case needs an exponential amount of space\footnote{For more details and/or a mathematically more precise formulation of this statement, see \cite{ModernCA}.}. On the other hand, in recent years there have been major improvements with regard to the run-time behaviour of the original Buchberger algorithm\footnote{See e.g.\ the results by J.-C.\ Faug\'ere in \cite{F4} and \cite{F5}.}, which have resulted in faster and thus more powerful algorithms to compute Gr\"obner bases. 

As we could see in the above examples, even moderately sized small circuits already give rise to mid-sized equations\footnote{Indeed we must not regard the results, which have been presented up until now, as {\sl large} -- the maximal computing time used in the above examples in the worst case, which was the calculation of all the solutions of the Chebyshev RC filter, was ten minutes.}, which again -- by way of the methods from computer algebra -- result in even bigger symbolic expressions for the sought-after design parameters. Thus the ``new toolbox'' computer algebra, the usage of which I try to recommend with this article, seems to be doomed from the start.

Still, bad behaviour {\sl in general} does not mean that the instruments from computer algebra {\sl always} behave \hbox{badly --} there might be classes of problems for which results are computed fast, or can be written down very economically. As electrical engineers know, there are close (but in the author's opinion by far not completely understood) connections between a circuit's topology and the algebraic structure of its network function, which may be taken advantage of with regard to design tasks.

Take on the other hand researchers in symbolic methods, who are accustomed with the phenomenon of swelling expressions arising in the analysis of circuits. As a workaround a number of them have proposed so-called ``sequences of expressions''\footnote{See \cite{RodanskiHassoun}.}. It is interesting to note that even electrical engineers without experience in methods of symbolic analysis may already be well versed in this ``sequential approach'', if only they knew classic filter theory, in particular the Cauer canonical forms of $LC$-circuits.

In form of an example, let us show, what we mean. Suppose we are given an $LC$-ladder network, consisting of three inductances and three capacitances, as shown in figure 5.  
\begin{figure}[!h]
\vskip -.4cm
\begin{center}
{\includegraphics[width=\linewidth,clip,keepaspectratio]{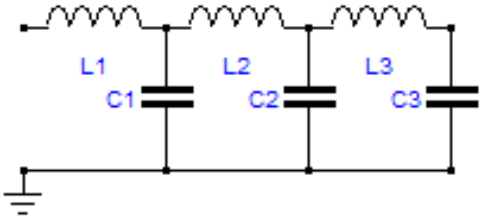}}
\vskip -1cm
\parbox{.6\linewidth}{
\caption{LC ladder in Cauer canonical form}
}
\end{center}
\end{figure}

As is well known its impedance $Z(s)$ can be described by a continued fraction
\begin{equation}
Z(s)=L_1\,s + \cfrac{1}{C_1\,s+\cfrac{1}{L_2\,s+\cfrac{1}{C_2\,s+\cfrac{1}{L_3\,s+\cfrac{1}{C_3\,s}}}}}.
\label{Frak}
\end{equation} 
That, in fact, is just short hand for a sequence of expressions. This can be seen by deducing (\ref{Frak}) with a process of alternatively combining impedances in series and in parallel. Thus we have
\begin{equation}
Z(s)=L_1\,s + Z_1(s),
\nonumber
\end{equation}
where $Z_1(s)$ denotes the impedance of the ladder circuit, with $L_1$ having been removed. Further:
\begin{equation}
\frac{1}{Z_1(s)}=\frac{1}{\frac{1}{C_1\,s}}+\frac{1}{Z_2(s)}=C_1\,s + \frac{1}{Z_2(s)},
\end{equation}
where $Z_2(s)$ is the impedance of the circuit, that is left, when $L_1$ and $C_1$ have been removed. Continuing like this, taking away one more circuit element in each step, this process can be repeated, until we get to one last impedance $Z_5(s)$ (which would just consist of $C_3$).

Reinserting the various $Z_i$ we get
\begin{eqnarray*}
Z(s)&=&L_1\,s + Z_1(s)\\
&=&L_1\,s + \frac{1}{C_1\,s + \frac{1}{Z_2(s)}}\\
&=&\dots,
\end{eqnarray*}
and thus are able to calculate $Z(s).$

When conversely one wants to realize an LC trans\-impedance function as an LC ladder, classic theory\footnote{See e.g.\ \cite{Weinberg}.} tells us to expand the given rational function into a continued fraction (an algorithm which in case of rational functions always comes to an end after a finite number of steps) and read off the element values directly from this expression\footnote{Again, nowadays a computer algebra system might be of help for calculating this expansion without much effort, see \cite{MapleBook}.}, which according to theorems in classic filter theory all are positive.

If we want to bring together the capabilities of a CAS for solving algebraic equations and the sequence of expression approach in our example, we first observe that we can easily calculate the form of the intermediate impedances $Z_i,$ i.e.\ we have
\begin{align}
Z_5(s) &= \frac{1}{C_3\,s} = \frac{1}{*s},\quad
Z_4(s) = L_3\, s + Z_5(s) = \frac{*s^2+1}{*s},\nonumber\\
\frac{1}{Z_3(s)} &= C_2\, s + \frac{1}{Z_4(s)} = \frac{*s^3+*s}{*s^2+1},\dots,\nonumber\\  
Z_1(s) &= \frac{a_4 s^4 + a_2 s^2 + 1}{a_5 s^5+a_3 s^3 + a_1 s},\quad
\hbox{and finally}\nonumber\\
Z(s)&= L_1\,s+Z_1(s)\nonumber\\
&=
\frac{(a_5\cdot L_1)s^6+(a_3\cdot L_1 + a_4)s^4+(a_1\cdot L_1 + a_2)s^2 + 1}
{a_5 s^5 + a_3 s^3 + a_1 s}\nonumber
\end{align}
with $*$ and the $a_i$ denoting polynomials in the circuit para\-meters $C_1,C_2,C_3,L_2,L_3.$ 

A sensible design task would now be to specify an impedance function
\begin{equation}
Z_{des}(s)=\frac{k\,(A_6 s^6 + A_4 s^4 + A_2 s^2 + 1)}{A_5 s^5 + A_3 s^3 + A_1 s},
\end{equation} 
with real numbers $k, A_1,\dots,A_6,$ 
and to demand $Z(s)=Z_{des}(s).$ 
Analogously to the above example of the BJT amplifier, by matching coefficients, this results in six nonlinear equations, which have to be solved for the variables $a_1,\dots,a_5$ and $L_1.$ Here we only present the (easy) solution (determined by Mathematica again) for the first step, which is
\vskip-1.5em
\begin{eqnarray*}
&& a_2= A_2-\frac{A_1\cdot A_6}{A_5},\quad a_4= A_4-\frac{A_3\cdot A_6}{A_5},\\
&& a_5 = \frac{A_5}{k},\quad a_1= \frac{A_1}{k},\quad a_3= \frac{A_3}{k},\\
&& L_1= \frac{A_6\cdot k}{A_5}.
\end{eqnarray*}
\noindent
To calculate further element values, this procedure needs to be repeated for the other $Z_i.$ Thus one has to solve next the equations resulting from $\frac{1}{Z_1(s)}=C_1\,s+\frac{1}{Z_2(s)},$ where $Z_1(s)$ should still be described in terms of the coefficients $a_i,$ and $Z_2(s)$ is given by the ansatz
\begin{equation}
Z_2(s)=\frac{b_4\,s^4+b_2\,s^2+1}{b_3\,s^3+b_1\,s}
\end{equation} 
with new unknowns $b_1,\dots,b_4.$
This has to be continued until we have equations for each set of coefficients of $Z_i$ in terms of the coefficients appearing in $Z_{i-1},$ plus additional equations for the element values. Starting from those equations, successive reinserting leads to the desired expressions for the element parameters. 

One has to concede that -- compared to the traditional continued fraction expansion -- this procedure looks very cumbersome. But in fact, if one looks a little bit closer, it is equivalent to repeatedly doing polynomial division as part of a continued fraction expansion. 
Furthermore our approach might be generalized to other impedance functions which need not come from ladders, but still ultimately result from a sequence of series, respectively parallel reductions applied to the initially given circuit, i.e.\ the process of replacing two impedances in series, or in parallel by a single equivalent impedance\footnote{However we need to be careful with the claim for generality, since if we combine e.g.\ two resistances $R_1,R_2$ in series to a single resistance $R,$ from the resulting equation $R=R_1+R_2,$ when only given the value of $R,$ we cannot deduce uniquely the values of the $R_i.$}. Beyond this there is a further generalization to handle Delta-Wye and Wye-Delta transformations. 

On the other hand by results from Epifanov and Truemper, it is known that every planar two terminal graph can be reduced to a single edge by these four kinds of transformations (plus removing loops and pendant edges). Moreover there is an algorithm, developed by Feo and Provan\footnote{As a reference for these results in graph theory, see the original paper \cite{FeoProvan}.}, which calculates the sequence of necessary graph transformations. Since to each of these transformations there corresponds an equation, which gives the value of the resulting element(s) in terms of the values of previously existing elements, there always is a {\sl sequence of equations} which leads from the initially given element values to the driving point impedance of a planar two terminal circuit\footnote{This algorithm has been successfully implemented in a student's minor thesis \cite{DomannStudien} at the TU Braunschweig in 2003.}.

Circuit design for (midsized to large) planar two terminal circuits, which consist of impedances only, thus becomes the task of ``reversing'' these equations, in an analogous manner to what we did with LC ladders.

 The author of this paper firmly believes, that computer algebra systems will be a valuable tool in tackling this particular problem. The challenge remains to find classes of circuits for which the {\sl sequence of reversed equations} can be effectively calculated, or to prove -- what might be also possible, and would be an interesting result in itself -- that, besides the classic, continued- or partial-fraction-topologies attached to the names of Cauer and Foster, no such further classes exist.  


\bibliographystyle{IEEE}



\section*{Note added to the Electronic Version}
In this electronic document keywords, ACM and MSC classifications have been added. \hfill (March~14th (Pi Day), 2009)

\begin{biography}
[{\includegraphics[width=1in,height=1.25in,clip,keepaspectratio]{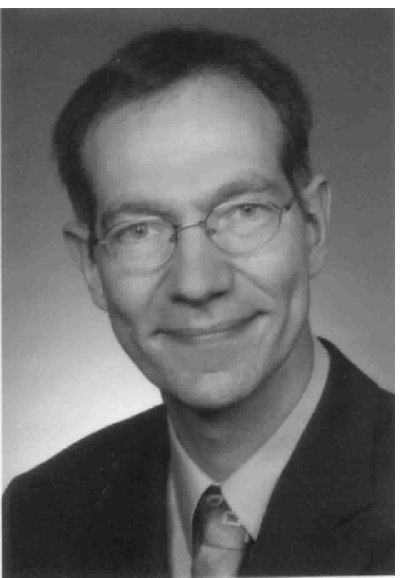}}]
{Eberhard H.-A.~Gerbracht}
received a Dipl.-Math.\ degree in mathematics, a Dipl.-Inform.\ degree in computer science, and a Ph.D. (Dr.\ rer.nat.) degree in mathematics from the Technical University Braunschweig, Germany, in 1990, 1993, and 1998, respectively.

\small From 1992 to 1997,  he worked as Research Fellow and Teaching Assistant at the Institute for Geometry at the TU Braunschweig. From 1997 to 2003 he was an Assistant Professor in the Department of Electrical Engineering and Information Technology at the TU Braunschweig. After a two-year stint as a mathematics and computer science teacher at a grammar school in Braunschweig and a vocational school in Gifhorn, Germany, he is currently working as tutor, and independent researcher in various areas of mathematics. His research interests include combinatorial algebra, C*-algebras, the history of mathematics in the 19th and early 20th century and applications of computer algebra and dynamical geometry to graph theory, calculus, and electrical engineering. 

\small
Dr.~Gerbracht is a member of the German Mathematical Society (DMV), the Society of Computer Science Teachers in Lower Saxony within the Gesellschaft f\"ur Informatik (GI-NILL), and founding member of the society ''Web Portal: History in Braunschweig - \href{http://www.gibs.info}{www.gibs.info}''.

\end{biography}

\end{document}